# Big Pharma, little science?
# A bibliometric perspective on big pharma's R&D decline[1]

Ismael Rafols[1,2,*], Michael M. Hopkins[1], Jarno Hoekman[3],
Josh Siepel[1], Alice O'Hare[1], Antonio Perianes-Rodríguez[4] and Paul Nightingale[1]

[1] SPRU – Science and Technology Policy Research, University of Sussex, Brighton, England
[2] INGENIO (CSIC-UPV), Universitat Politècnica de València, València, Spain
[3] Department of Clinical Pharmacology, University Medical Center Groningen, University of Groningen, the Netherlands
[4] Carlos III University of Madrid, Department of Library and Information Science, SCImago Research Group, Madrid, Spain

*Corresponding author: i.rafols@sussex.ac.uk, SPRU, University of Sussex, Freeman Centre, Brighton, BN1 9QE, England. Tel. +44-1273-873087.

**Abstract**

There is a widespread perception that pharmaceutical R&D is facing a productivity crisis characterised by stagnation in the numbers of new drug approvals in the face of increasing R&D costs. This study explores pharmaceutical R&D dynamics by examining the publication activities of all R&D laboratories of the major European and US pharmaceutical firms during the period 1995-2009. The empirical findings present an industry in transformation. In the first place, we observe a decline of the total number of publications by large firms. Second, we show a relative increase of their external collaborations suggesting a tendency to outsource, and a diversification of the disciplinary base, in particular towards computation, health services and more clinical approaches. Also evident is a more pronounced decline in publications by both R&D laboratories located in Europe and by firms with European headquarters. Finally, while publications by Big Pharma in emerging economies sharply increase, they remain extremely low compared with those in developed countries. In summary, the trend in this transformation is one of a gradual decrease in internal research efforts and increasing reliance on external research. These empirical insights support the view that large pharmaceutical firms are increasingly becoming 'networks integrators' rather than the prime locus of drug discovery.

**Keywords:** Pharmaceuticals, bibliometrics, outsourcing, Europe, globalisation, research network, innovation, collaboration

**Highlights**

- We analyse the publications by R&D laboratories of the top 15 pharmaceutical firms.
- We observe a slow decline in their total number of publications and field share.
- A more pronounced decline in publications by R&D laboratories located in Europe.
- There is more external collaboration and research in non-traditional disciplines.
- The results suggest that Big Pharma firms are increasingly network integrators.

---

[1] This article was nominated to Best Paper Award in the 2012 DRUID Conference (http://www.druid.dk).

## 1. Introduction

Studies from a variety of sources and perspectives suggest that the pharmaceutical industry is facing a productivity crisis and is undergoing a substantial transformation. A stagnant or declining number of new chemical entities (NCE) are approved by regulators such as the US Food and Drug Administration (FDA) each year in spite of sustained, major increases of R&D expenditures (Munos, 2009; Pammolli et al., 2011). In response, pharmaceutical firms have engaged in a series of major mergers with each other; undertaken waves of acquisitions involving smaller drug discovery firms; closed R&D sites, particularly in Europe and the US; sought cost savings through rationalisation (LaMattina 2011) and opened R&D laboratories in emerging countries with large markets such as India and China (Anon 2010). The industry is increasingly outsourcing R&D to external research organisations, which is perceived to improve efficiency (Baum 2010). Governments are supporting these trends by increasingly focusing public sector funding on 'translational' research (Collins 2011).

In this paper we explore these shifts by studying changes in the publication activities of the world's 15 largest pharmaceutical firms in Europe and the US ('Big Pharma'). Historically, the role of Big Pharma in the production of scientific research has been important, with a collective R&D investment of ~£46bn per annum between 2004-2007 (see Table 1) compared to an average annual spend of £15.25 by the US National Institute of Health (NIH) over the same period.[2] As a result, the publication activity of these pharmaceutical firms is large (~10,000 document/year) and may be explored to trace changes in the industry. Publication data does not provide direct insight into the R&D processes because a variety of factors shape and mediate propensity to publish, but as discussed in the next section when handled with care and contrasted with other data sources, publication rate and patterns may shed some light on dynamics in the quantity, areas and modalities of R&D efforts, and thus provide glimpses of the underlying processes of change processes of change (Tijssen, 2004, pp. 713–715).

We will explore shifts in Big Pharma activity from three perspectives: How has the knowledge base of pharma changed? How has its organisational structure evolved? How have pharmaceutical firms relocated their R&D activities?

First, in the cognitive sphere, the shift since the 1970-80s with the advent of biotechnology from a random-screening regime, towards a guided-search regime could be expected to make pharma R&D more reliant on basic biological research (McKelvey et al., 2004; Hopkins et al., 2007). However, the cumulative nature of competence acquisition by firms means that this process would be expected to occur in a progressive manner (Nightingale and Martin, 2004; Hopkins et al., 2007), rather than in a disruptive way (Hwang and Christensen, 2008). What do publications data tell us about the changes in the knowledge base?

Second, at the organisational level, the interaction between emerging 'biotech' firms and pharmaceutical industry has been presented as the prototypical example of the organisational network as the 'locus of innovation' (Powell et al., 1996), with an estimate by UK drug developers that they outsource 25% of their R&D (Howells et al., 2008). Recent empirical research has also shown evidence for this shift in drug discovery. Munos (Munos, 2009, p. 965) showed that the share of the drugs approved by large firms in the US has steadily decreased from ~80% in 1980 to about ~50% in 2008. Kneller (Kneller, 2010) reported that at least half of the new drugs discovered in the USA between 1998 and 2007 originated from public laboratories or small firms (see also (Angell, 2004, pp. 52–73)).[3] If this is the case, can we see an increase in the dependence of Big Pharma on external collaborations? Is there evidence of pharmaceutical R&D becoming less 'R' (research) and more 'D' (development)?

---

[2] The average per year is $28.5Bn - See NIH budget at http://www.nih.gov/about/director/budgetrequest/
[3] In the case of biotechnology-related patents the growth of small firms is even faster: from less than 15% in 1990-94 to more than 35% in 2000-04, with big pharma shrinking from ~42% to ~31% (Patel et al., 2008).

Third, from a geographical perspective, the internationalisation of R&D either by off-shoring internal R&D or by outsourcing it using external collaborations, has received significant policy and media attention. The perceived relocation of pharmaceutical activities is often accompanied by stories of European weakness in comparison to the US (Tijssen, 2009). Given the relative importance of the pharmaceutical industry in various European economies, and its position as one of the (few) high-tech industries with a dominant European base, the potential decline of this sector has been a regular concern in the EU (e.g. see (Tijssen, 2009; Pammolli et al., 2011)). Do publications support the view of a European decline? Given the observation that interactions between publicly funded research and firms' R&D efforts often take place in close geographical proximity, is this European decline also visible in Big Pharma's collaborative links?

**Table 1. List of European and USA pharmaceutical firms included in this study.**

| Firm | Publications 1995-2009 | Employees (2008) | Mean R&D spent/year (2004-2007 in M £) |
|---|---|---|---|
| **GlaxoSmithKline** | 19,331 | 101,133 | 3,186 |
| **Novartis** | 15,477 | 96,717 | 3,604 |
| **Hoffmann–La Roche** | 14,351 | 80,080 | 4,195 |
| **AstraZeneca** | 11,378 | 66,100 | 2,740 |
| **Sanofi-Aventis** | 11,211 | 98,213 | 3,722 |
| **Bayer** | 8,125 | 107,299 | 2,270 |
| **Novo Nordisk** | 3,378 | 31,062 | 837 |
| **Boehringer Ingelheim** | 3,036 | 41,300 | 1,425 |
| **Aggregate EU** | **84,863** | **621,904** | **21,979** |
| **Pfizer** | 23,290 | 129,226 | 7,371 |
| **Merck** | 21,697 | 106,200 | 4,540 |
| **Eli Lilly** | 9,584 | 40,500 | 2,144 |
| **Johnson & Johnson** | 7,197 | 118,700 | 4,576 |
| **Abbott** | 6,482 | 69,000 | 1,440 |
| **Bristol-Myers Squibb** | 6,349 | 35,000 | 2,016 |
| **Amgen** | 5,070 | 16,900 | 1,908 |
| **Aggregate USA** | **78,194** | **515,526** | **23,995** |

Note: Publications include those of subsidiaries, acquisitions and parent firms of mergers. Source: 2009 UK R&D Scoreboard (BIS, 2009)

There is a lack of agreement among scholars about what R&D off-shoring to developing countries precisely entails (Ujjual et al., 2011). Due to the relatively distinct activities pursued in the stage of drug discovery (lab based) and drug development (clinically based) it may well be the case that R&D off-shoring takes place for some activities and not for others. Some scholars argue in this respect that there is a shift towards a globalisation of innovation, for example via externalisation of clinical trials to Contract Research Organisations (CROs) (Archibugi and Iammarino, 1999), whereas others suggest that local R&D centres are concerned with adaptation to local markets, for example by focusing on research on diseases that are prevalent in tropical areas or among certain population groups (von Zedtwitz and Gassmann, 2002). What do publications tell us about these alleged geographical shifts?

To our knowledge, this is the first large scale study of pharmaceutical publication trends. The only studies published previously are by Robert Tijssen, who showed trends of collaborative modes (Tijssen, 2004) or focused on the location of pharmaceutical R&D of European firms (Tijssen, 2009). Our analysis covers the world's largest pharmaceutical firms and uses novel visualisation tools to intuitively convey to a non-expert audience the knowledge base of these firms (Rafols et al., 2010) and their collaboration networks (Perianes-Rodríguez et al., 2010). Data and bibliometric visualisations are made available in the Complementary Files listed in Appendix 1 and in a dedicated website.[4]

Publications give only a partial view of a firm's activities, and section 2 discusses the limitations of such a focus before describing in Section 3 the materials and methods employed to track the publications of large pharma. Section 4 provides a detailed analysis of publication output exploring each of the three areas outlined above. Section 5 provides discussion and conclusions where we emphasise that our empirical insights support the prevalent view that large pharmaceutical firms are increasingly becoming 'networks integrators' rather than the prime locus of drug discovery (Hirschler and Kelland 2010, Hopkins et al. 2007).

**2. The limitations of using publication data to track the R&D activity of Big Pharma**

Publications cannot be assumed to be a reliable proxy to describe the dynamics of research in a private firm, even in a science-based area such as pharmaceuticals. For example, the concerns raised over the scientific integrity of research conducted by pharmaceutical companies are likely to have affected publication strategies, especially in the clinical fields (e.g. (Angell, 2004; Smith, 2005)). As well as a decline in R&D, we also observe that pharmaceutical companies may face decreased legitimacy in science which may be reflected in their publishing patterns (Sismondo, 2009) (Sismondo 2009). Careful analysis is thus needed to make inferences from publication data.

To understand how Big Pharma's R&D activity differs from its publication activity, it is helpful to consider that publication activity has traditionally been associated with *Open Science* institutions. By Open Science we mean here a distinct organizational sphere where rapid disclosure of new research results and sharing of associated methods and materials is encouraged (David, 1998). Open Science should not be confused with *Open Innovation* (Chesbrough, 2003), a model in which firms develop innovation by sharing knowledge with a variety of non-profit and for-profit organisations –but without the knowledge being necessarily in the public domain. Open Science is based on the pursuit of priority, for example to claim credit for discovery and to hasten diffusion of knowledge, and as such encourages the rapid disclosure of research findings in scientific journals (Merton, 1973; Stephan, 1996). In contrast, firms often rely on secrecy and protective mechanisms such as patents to limit knowledge spillover risks and ensure returns to their investments (Dasgupta and David, 1994), even if, they operate in an Open Innovation model. This implies that the contributions of Big Pharma to Open Science cannot be simply considered an unbiased reflection of their research efforts and their scientific discoveries. Rather, scientific publication activity in firms is not only driven by the pursuit of scientific reward but also by commercial interests and by pressures imposed on firms by prescribers, healthcare payers and regulators to disclose data.

In the case of pharmaceutical research it is important to distinguish between the motivations firms have to publish during drug discovery and during drug development. With respect to the former, Hicks suggested that firms publish in order to 'participate in the barter-governed exchange of scientific and technical knowledge' and to send signals to investors (Hicks, 1995, p. 421). Adopting an Open Science strategy is in this case considered necessary in order to connect to the scientific community and to access its resources in the form of knowledge, qualified labour and informal advice (Hicks, 1995; Cockburn and Henderson, 1998). Investments in R&D are in this context a means to create absorptive capacity which is necessary both to take advantage of (upstream) research conducted outside the organizational boundaries of pharmaceutical firms. The advent of biotechnology has been associated with an increase of explicit interaction between industry and academic science (McKelvey et al., 2004). Firm strategies have thus moved towards a more networked and Open Innovation model (Powell et al., 1996; Chesbrough, 2003), in which publishing is seen as positively associated with innovative success (Jong and Slavcheva, 2012). However whether Open Innovation leads to more Open Science is an open question. Indeed our results suggest the reverse.

---

[4] www.interdisciplinaryscience.net/pharma

In drug development, it can be argued that incentives to publish are higher than in other science-based industries, due to the highly regulated nature of drug development and the importance of clinical evidence for user uptake of innovations. Moreover, intellectual property will already be in place on the underlying compound before drug development, which limits the risks of free riders. Scientific publications are therefore primarily written to diffuse information about the effectiveness and safety of pharmaceuticals to a wide range of stakeholders. This process is especially incentivized in an evidence-based medicine paradigm (Timmermans and Berg, 2003; Montori and Guyatt, 2008) which is an attempt to ground medical decisions directly in available scientific evidence (e.g. clinical trials, systematic reviews). It follows that pharmaceutical firms make use of publications as marketing tools that need to be carefully constructed and employed in order to win support in regulatory or policy arenas (e.g. in to gain approval of clinical trials) and in clinical settings (e.g. credibility among doctors) (Smith, 2005; Sismondo, 2009). In drug development scientific publications are also used as competitive devices to promote the superiority of a compound vis-a-vis potential substitutes introduced by competitors (Polidoro and Theeke, 2011).

While publications are therefore important in drug discovery and drug development, their analysis needs to be undertaken carefully because publications serve a variety of purposes. Thus, changes in publication trends can reflect different underlying phenomena. For example, the propensity to publish may change when a firm shifts towards more science-based areas and wants to engage with academics (Hicks, 1995). Similarly, in drug development, publication may underestimate research in pharmaceutical firms if industrial scientists publish their own research using 'ghost writers', i.e. hidden behind the alleged authorship of academics (Sismondo, 2007).[5] Moreover, some portion of research may be expected to be held back from publication for reasons of commercial secrecy, as discussed above.

A final consideration is the extent to which tracking the publications of the 15 largest firms in an industry tells us about the industry as a whole. There are thousands of firms now engaged in pharmaceutical R&D. In comparison to Big Pharma these firms are either smaller or younger and some of them are engaged in non-traditional activities, for example biotech firms focusing on novel therapeutic modalities like cell therapies or RNA interference. While the sample here does not represent the industry as a whole, the firms tracked in this sample account for over 50% of the pharmaceuticals brought to market since 1950 (based on (Munos, 2009)) and hence represent the core of the traditional pharmaceutical industry. Furthermore because the publications tracked include those of firms acquired by Big Pharma over the studied period, the publications also incorporate the attempts by these firms to renew their capabilities.

Given all these limitations, we conclude that the analysis of publications *does not* in itself reflect the dynamics of Big Pharma's R&D. However, at the high level of aggregation we conduct this study (based on about 10,000 publications per year in total, with around 150 to 1,500 publications per firm annually) it does raise interesting questions on R&D trends and firm strategies which then can be discussed in light of complementary quantitative evidence such as the trends revealed in studies using a variety of other metrics such as patents and pharmaceutical projects (Kneller, 2010; Pammolli et al., 2011).

---

[5] This occurs because of the higher credibility of academic (perceived as 'disinterested') researchers' findings to investors and medical practitioners.

## 3. Materials and Methods

The papers authored by staff at 15 major European and USA pharmaceutical firms were downloaded from Thomson-Reuter's Web of Science (WoS). [Complementary File 1](#) (URL embedded) lists the firm names used in retrieval, including subsidiaries, and merger and acquisition targets. Information on each firm's acquisitions, mergers and subsidiaries was collected from their annual reports and the Recombinant capital database (http://www.recap.com/). The document types chosen were 'article', 'letter', 'note', 'proceeding paper', and 'review' for the period 1995-2009. A total of 160,841 records were obtained, standardised, processed and analysed with VantagePoint software (http://thevantagepoint.com/). Publications were classified as European when the affiliation contained at least one country in the European Union (EU) or in Switzerland, Norway, Lichtenstein and Iceland, i.e. the members of the European Free Trade Association (EFTA).

The freeware programmes *Pajek* (http://pajek.imfm.si/) and *VOSviewer* (http://www.vosviewer.com/, (van Eck and Waltman, 2010)) were used for visualisation. A description of the methods used for maps is presented in Appendix 2.

To compare our dataset of the publications of the top 15 pharmaceutical firms to the field of sciences relevant to pharmaceutical R&D in general, we made two baseline datasets. First, to look into number of organisations and authors per paper, publications of the top 200 journals in which pharma published were downloaded from WoS. Two thousand publications were randomly selected from these journals for the years 1995, 1999, 2004 and 2009 and this data was used as a baseline for Figure 5. Secondly, to investigate the relative change in the number of publications that list Big Pharma staff as authors per journal and field, the total number of publications for the period 1995-2009 of the top 350 Journals in which Big Pharma published were downloaded from the Journal Citations Reports. This data was used as a baseline for Table 2.

## 4. Results: Shrinking knowledge production

Big Pharma has published over 10,000 publications per year in 1995-2009. This is a substantial contribution in the biomedical area, equals to about 4% of all publications in their field (estimated from the 350 journals where Big Pharma has the most publications).

The first insight from this study is that Big Pharma have reduced the number of publications they produce by around 0.8% per annum when taking into account additional boosts to publication counts from subsidiaries and acquisitions prior to them joining their current parent. This amounts to a 9% decrease over 15 years, as shown in Figure 1. The results are a conservative estimate for the decrease, given that we are using full counting (i.e. without assigning fractions to co-authoring organisations) and collaborations are increasing over time (as discussed in section 4.2).

This decrease is in stark contrast with R&D expenditure by large firms in the industry, which have increased in the order of 50% to 400% a decade (Arrowsmith, 2012, p. 18) and the general inflationary tendency in publication volumes revealed in bibliometric studies of global scientific output (Persson et al., 2004; Leydesdorff and Wagner, 2009). In the areas where Big Pharma is publishing the most, we have estimated an annual growth rate of 1.1%, amounting to a 16% growth over the period (again, estimated from the 350 journals where Big Pharma has the most publications). As a result, publications by Big Pharma show a relative decrease, from 5% of the total in the specialised pharma fields in 1995 to 4% in 2009. This slow decline is consistent with other studies apparently showing an absolute decline of patenting in pharmaceutical US patent classes 424 and 514 (Subramanian et al., 2011, p. 68) and relative decline of patenting by big pharma in comparison to small firms in biotechnology (Patel et al., 2008, p. 51).

However, if one looks at the publications by the core firms in our sample (defined as those with the name of the parent firm or of mergers), one observes a modest increase of 0.6%, totalling an 8% growth over the period. The difference between growth in publications by core firms and a decrease in all Big Pharma publications can be attributed partly to R&D outsourcing, and partly to closure, absorption and often rationalisation into the core firm of the R&D laboratories of the acquired firm. For example the former laboratories of firms such as Searle, Upjohn and Warner-Lambert were closed some time after their acquisition by Pfizer (LaMattina, 2011). Merck is estimated to have reduced its workforce by 30% after its merger with Schering-Plough in 2009 (McBride and Hollmer, 2012). All the pharmaceutical industry (not just Big Pharma) is estimated to have reduced its workforce by 300,000 people since 2000 (Herper, 2011).

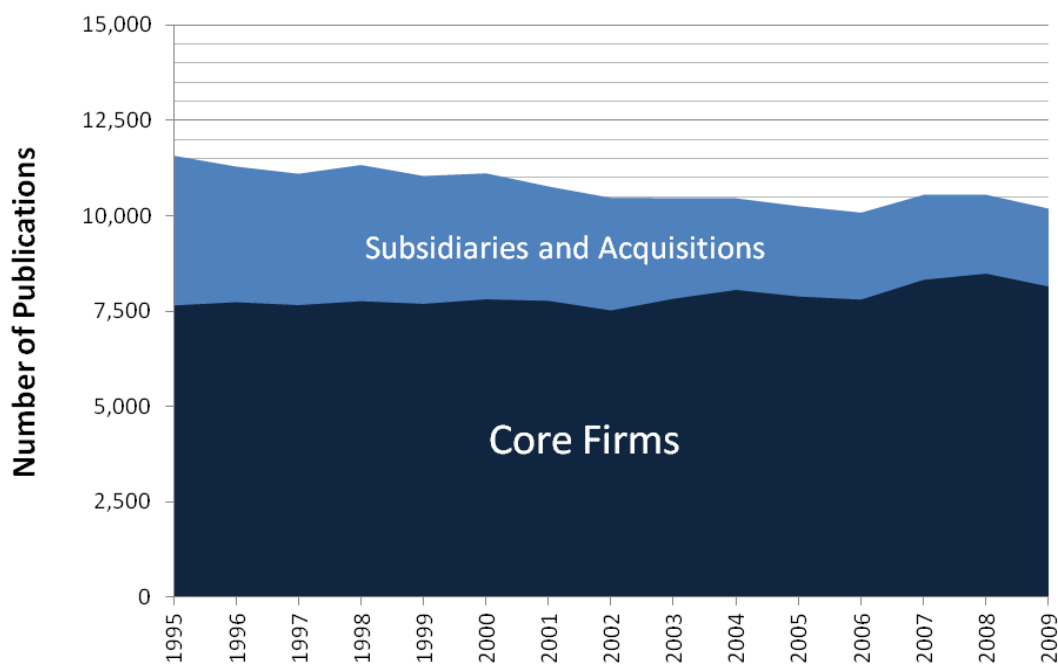

**Figure 1. Total number of publications per year of top 15 pharmaceutical firms.** 'Core firms' includes only publications by R&D labs of the main firm or (or antecedents' names) before a merger. 'Subsidiaries and Acquisitions' includes the outputs published under the name of the acquired firms before and after the transaction.

Figure 2 illustrates the wider dynamics of apparent core firm enlargement with overall R&D shrinkage in the case of the GlaxoSmithKline, which saw a sharp decrease in its aggregate number of publications in the aftermath of the merger of Glaxo-Wellcome and SmithKline Beecham in 2000. In the cases of Pfizer, Novartis and Bayer, a similar dynamic is observed for acquisitions. The only firm that shows significant growth in publications over the period is Johnson & Johnson. Data for each firm is available at [Complementary File 2](Complementary File 2).

One can assume that the firms showing lower number of publications in relation to their R&D investment are also those with the highest degree of R&D outsourcing. For example, Sanofi-Aventis, who has reduced its publications per year from ~1200 to ~400, report that 'in Feb 2011, 64% of our development portfolio consisted of projects originated by external R&D' (Sanofi-Aventis, 2010, p. 17). AstraZeneca states that 'We intend to increase our externalisation efforts to access the best, most cutting edge science, whatever its origin, with a target of 40% of our pipeline sources from outside our laboratories by 2014' (AstraZeneca, 2010, p. 29).. It therefore appears that the outsourcing strategy is very actively being pursued by Big Pharma.

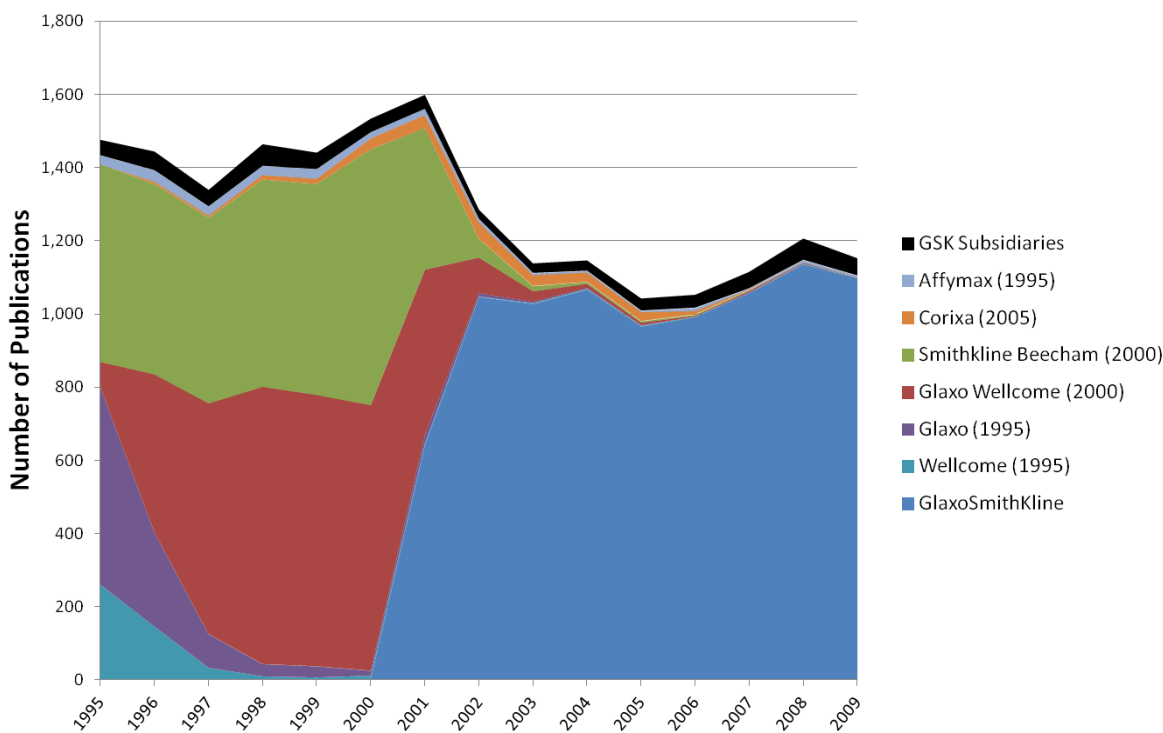

**Figure 2. Number of publications per year of GlaxoSmithKline, its subsidiaries, and the parent firms before merger.** Year of merger or acquisition shown in brackets. See other firms in Complementary File 2.

### 4.1 Cognitive shifts: Diversification of the knowledge base

Next we examine the areas of research where Big Pharma is most active and how they changed over time. Figure 3 shows the distribution of the publications by Big Pharma over the global map of science (Rafols et al., 2010).[6] Each node represents a research field as defined by Web of Science Categories. The position and colours of the nodes is given by their relative similarity (strong similarity is shown with links). The size (area) of nodes shows the percentage of publications in that area. Figure 3 illustrates that the knowledge base of Big Pharma is centred on biomedical sciences (green nodes: Pharmacology and Pharmacy, Biochemistry and Molecular Biology), with some important contributions in Chemistry (in blue: Organic Chemistry and Medicinal Chemistry), Immunology and Infectious Diseases, and then a few smaller areas of Clinical Medicine (in red). Data is available at Complementary File 3, for the aggregate and for each firm.

Figure 4 shows the areas with positive (top) and negative (bottom) growth in publication numbers. The visualisation reveals a broad pattern of diversification, with a decrease in the traditional yet still dominant biomedical and chemical disciplines and an increase in peripheral areas related to new techniques (e.g. computational biology and related), and disciplines more oriented to clinical applications of therapeutics or health services.[7] The latter observation is consistent with the observation that 'industry has shifted resources away from drug discovery to late clinical developments' (Munos and Chin, 2011, p. 1). For example, the proportion of pharma-biotech alliances in the pre-clinical stage decreased from 46% to 38% in 2007-2011, while those in marketed stage increased from 24% to 28% of the total (Ratner, 2012, p. 119).

In the cognitive sciences, one also observes a move away from basic science (e.g. Neuroscience) and towards more applied fields (Psychiatry and Clinical Neurology). The decrease in publications related to plants and environment is possibly due to the externalisation of the agrochemical divisions from pharmaceutical firms. For example, Syngenta was created in 2000 when Zeneca and Novartis span out and merged their agrochemicals divisions.

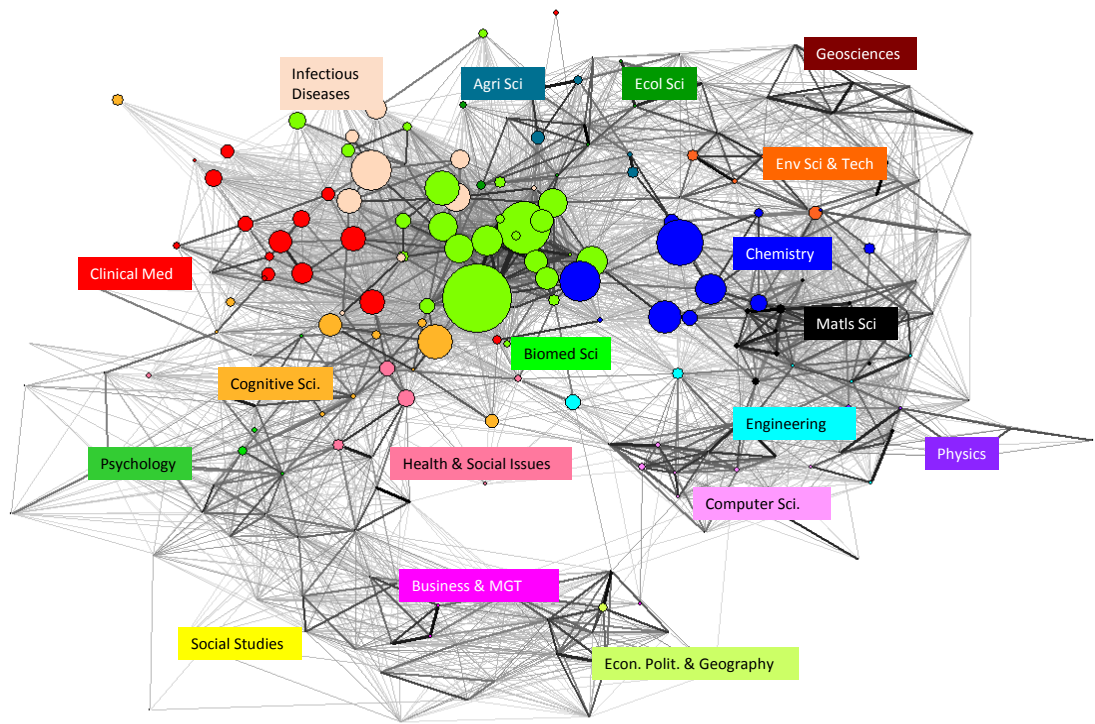

**Figure 3. Distribution of publications by the top 15 pharmaceutical firms over the global map of science.** Nodes represent subject categories, with the area proportional to number of publications. The position of nodes is determined by similarity in citation patterns between the subject categories they represent.

The growth of clinical areas such as oncology or rheumatology are consistent with data from pharmaceutical project data (Pammolli et al., 2011, p. 431). However, for other areas, such as cardiovascular, research output in terms of publications is slowly increasing (2.5% over 10 years) in spite of a substantial (-5%) decrease in the number of projects (Pammolli et al., 2011). Such apparent inconsistencies, though not surprising given variations in field definition and unit of analysis, warn against hasty interpretations of data.

---

[7] This diversification is captured by an increase in the Shannon-Wiener diversity (from 3.87 to 3.95).

We also tried to visualise the specialisations of each firm using Web of Science publication data, but found Big Pharma had quite similar profiles at this coarse level of aggregation.[8] In order to improve the granularity of description, we created a map based on the 191 journals that where Big Pharma publishes most frequently, available in [Complementary File 4](). See also interactive website: http://www.scimago.es/perianes/spru/Interaction_Companies.html.

The results of the visualisation were mixed. These more granular maps (not shown in printed version) capture some of the specialisation patterns: for example, Eli Lilly's focus on mental illness, or Novo Nordisk's effort in endocrinology. However, the maps were less successful at showing the differences in focus between larger firms such as Pfizer, Novartis or Merck. This may suggest isomorphic pressures in the industry as leading firms are joined by fast followers into new therapeutic areas (DiMasi and Faden, 2011), which is quite likely given the industry's reputation for producing 'me-too' drugs (Angell, 2004, pp. 74–93)). However, we believe that this might also signal the need for new analyses requiring more sophisticated approaches, for example using keywords such as the Medical Subject Heading (MeSH) provided by PubMed (Leydesdorff et al., 2012).

The overall picture is that Big Pharma's is reducing research in areas of traditional core expertise, namely in biomedical and chemistry areas, while at the same time diversifying into areas such as computational biology and fields that are closer to the patient (or perhaps market) such as health services and clinical research. However, as Big Pharma moves into new fields it relies more on external collaborations, as discussed in the next section.

---

[8] A previous study had shown no effect of these small field differences between firms on innovation performance (D'Este, 2005, p. 37).

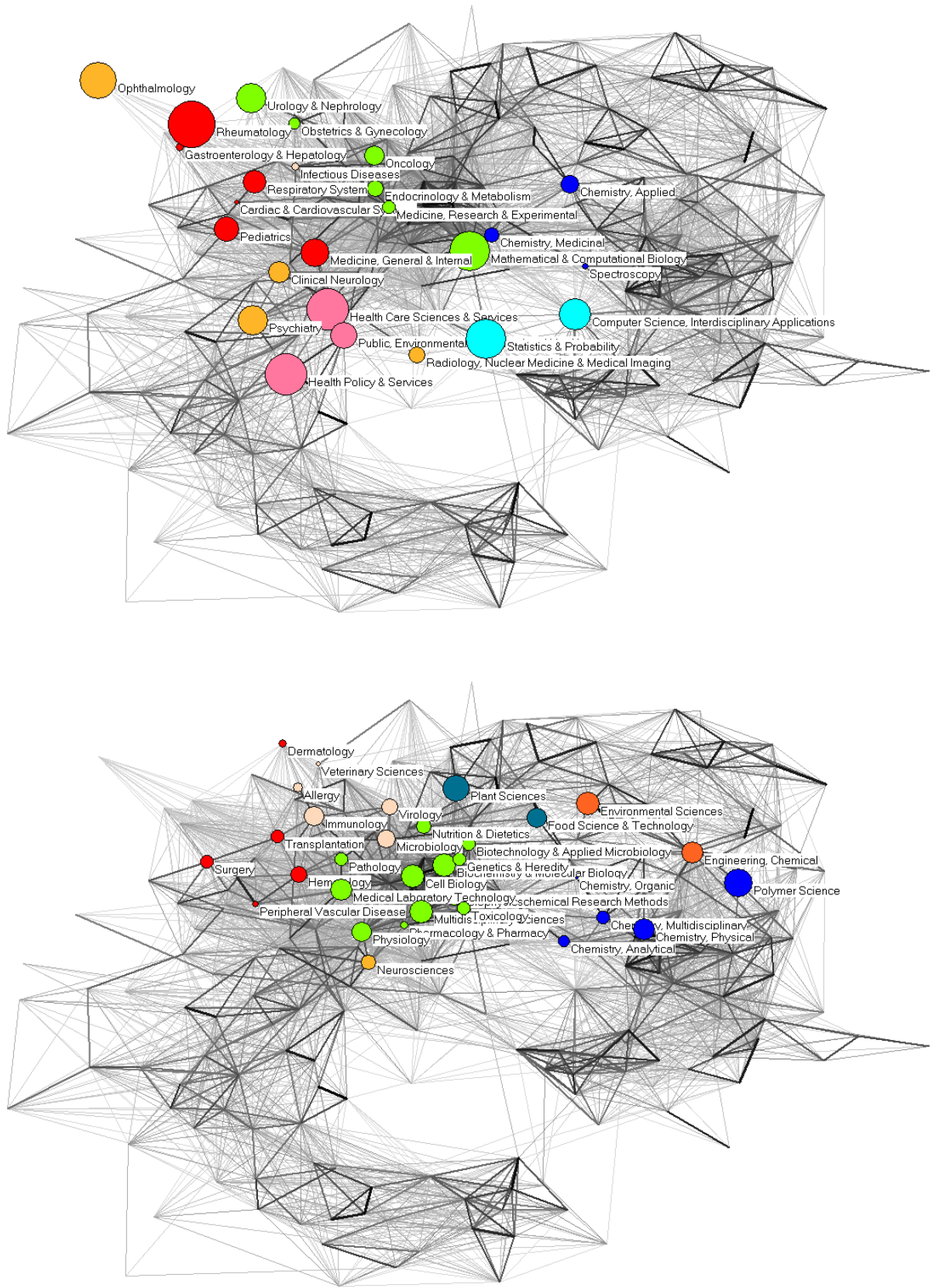

**Figure 4. Growth rate (top) and decrease rate (bottom) of publications by pharmaceutical firm in different scientific subject categories**. Area of nodes is proportional to the growth (decrease) in publications. Only subject categories with more than 0.5% of the total publications are shown.

## 4.2 Organisational shifts: Increasing collaborations and out-sourcing

The number of collaboration in science has been steadily increasing for decades (Hicks and Katz, 1996; Wuchty et al., 2007). This trends towards 'team science' has been particularly pronounced in the biological sciences, medicine and neuroscience (Porter and Rafols, 2009, p. 730). Our data shows that such trends are even stronger in Big Pharma, but there is a twist: although Big Pharma is collaborating more, it is their partners that are more often taking the intellectual lead in the work that follows from such collaborations.

Figure 5 illustrates the increase of the number of authors and the number of organisations per article for pharmaceuticals in comparison with a sample of papers (baseline) in related fields (2,000 publications per year of analysis randomly extracted from the 200 top journals of pharma publication).[9] Not only are pharmaceutical papers more collaborative, but their co-authorships are growing faster than average in the fields in which they publish.

The extent to which Big Pharma collaborations are increasingly led by the external partners is shown in Table 2. There is a significant decrease in the percentage of Big Pharma-based first authors in collaborative publications (from 43% to 35%), and of Big Pharma-based corresponding authors (from 41% to 34%) over the period studied.

This trend is more prominent in the fields and journals into which Big Pharma has more recently entered (arguably those where new competences are more likely to be demanded), in contrast to the fields and journals where Big Pharma is shrinking its publication output (which are those of traditional core expertise). In the fields with increasing output, the share of Big Pharma corresponding authors decreased from 44%- to 36% in 1998-2009 (and from 47% to 39% in the ten fastest growing). In the case of areas with declining output, the share of pharma's corresponding authorship remained quite stable, declining only from 52% to 49% (from 52% to 48% in those declining fastest). Likewise, the percentage of corresponding authors in growing journals decreased from 52% to 47% (from 70% to 60% in ten fastest growing), whereas the percentage in shrinking journals only declined marginally from 51% to 49% (from 51.2% to 50.5% in ten fastest shrinking). The inertia in the maintenance of the core expertise and the difficulty in catching up with new fields, is a general characteristic of firm dynamics given the cumulative nature of firms technological competences (Patel and Pavitt, 1997; Hopkins et al., 2007).

The overall picture is consistent with a trend towards R&D outsourcing, driven by pressures to seek cost efficiencies (Baum, 2010). Outsourcing entails a reduction of internal R&D laboratory capacity and the parallel expansion of alliance networks with small firms and academic laboratories (Powell et al., 1996), in which Big Pharma is an intellectual follower rather than a leader. This does not necessarily imply that Big Pharma's ability to capture value from these alliances is reduced – therefore, the implications of this shift will need further study.

---

[9] The increase in collaborations occurs both in 'small' and 'big science': a doubling or trebling in collaborations is observed both for very big teams (6 to 10 organisations, or 20 to 40 authors), and for large-scale projects (>20 organisations and >40 authors).

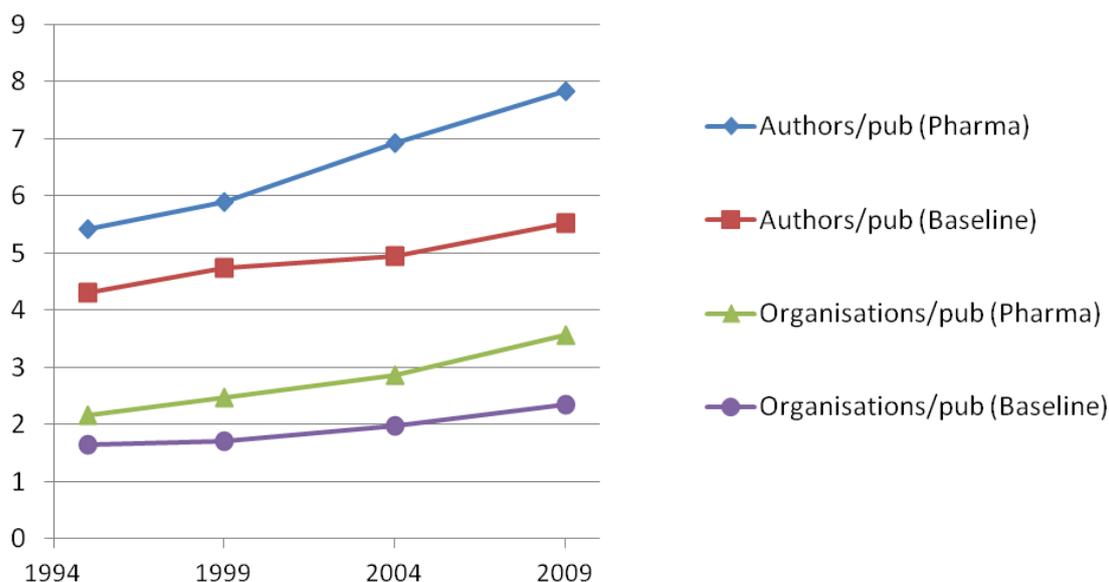

**Figure 5. Trends in collaborative practices in pharma R&D publications, compared to control publications in the same scientific fields (baseline)**.

**Table 2. Trends in scientific leadership in co-authored publications**

| Year | 1998[1] | 2002 | 2006 | 2009 |
|---|---|---|---|---|
| **% publications with external collaboration** | 62.1% | 67.6% | 69.1% | 71.7% |
| % organisations collab. baseline (estimate) | 55.0% | 59.0% | 63.2% | 66.4% |
| | | | | |
| **Pharma-based first author** | 43.1% | 38.4% | 36.2% | 35.0% |
| | | | | |
| **Pharma-based corresponding author** | 40.7% | 36.6% | 35.0% | 33.6% |
| In expanding fields[2] | 43.9% | 38.9% | 35.9% | 35.6% |
| In shrinking fields[2] | 52.3% | 50.1% | 50.5% | 48.8% |
| | | | | |
| **Pharma corresp. author (top 350 journals)** | 50.8% | 46.5% | 47.3% | 47.0% |
| In expanding journals | 52.4% | 48.0% | 48.1% | 46.6% |
| In shrinking journals | 50.9% | 46.4% | 47.6% | 48.8% |

Notes: (1) Data starts in 1998 due to partial missing data in WoS on corresponding and first authors in 1995-1997. (2) Since some journals are assigned to various fields (Web of Science Categories), this figure is a slight overestimate.

## 4.3 Geographical shifts

### 4.3.1 The decline of Big Pharma's European R&D

It has been suggested that European pharmaceutical firms have weakened over recent decades (McKelvey et al., 2004). The trends confirm such decline both in terms of the publications of all European R&D laboratories, shown in Figure 6, left, including European labs of American firms, such as Pfizer's site at Sandwich, UK (which has closed since this data was collected, illustrating the point), and of all laboratories of firms with European headquarters (shown in Figure 6, right, including USA labs of European firms). Beyond the decline in number of publications, European-headquartered firms have also decreased the proportion of publications that they lead, as shown by number of corresponding authors, from 45% to 35% (Figure 6, right), whereas US-based firms have undergone a less marked decline (from 52% to 47%). Geographical data discussed in this section is available in Complementary File 5.

Figure 7 shows that the decline of Big Pharma's European publications has been concentrated in the UK, Switzerland and France. In contrast, Big Pharma firms have maintained their German publication output and increased their output in smaller European countries such as Sweden, Belgium and Denmark.

Part of the European decline may be attributed to the fact that whereas Europe-based firms have located an important share of their R&D activities in USA laboratories (producing about 35% of their publications), USA-based firms on the other hand have a smaller presence in Europe (producing only 22% of their publications). This is illustrated in Figure 8 (left).

Such imbalances in the USA versus European presence of Big Pharma has been previously well documented, e.g. by studies on publications (Tijssen, 2009) and patent inventors (Friedman, 2010; Pammolli et al., 2011). In the latter case, patent concentration in the home continent of the firm is even more acute, as shown in Table 3. These data should be taken as conservatives estimates of home country bias because in this study publication data counts the publications of acquired firms with the parent firm –which means that acquisitions explain a very important proportion of the non-home publications of firms. For example, Roche's large percentage of USA publications is partly explained by its acquisition of Genentech; or Johnson and Johnson's large European presence is largely due to its Belgian subsidiary Janssen.

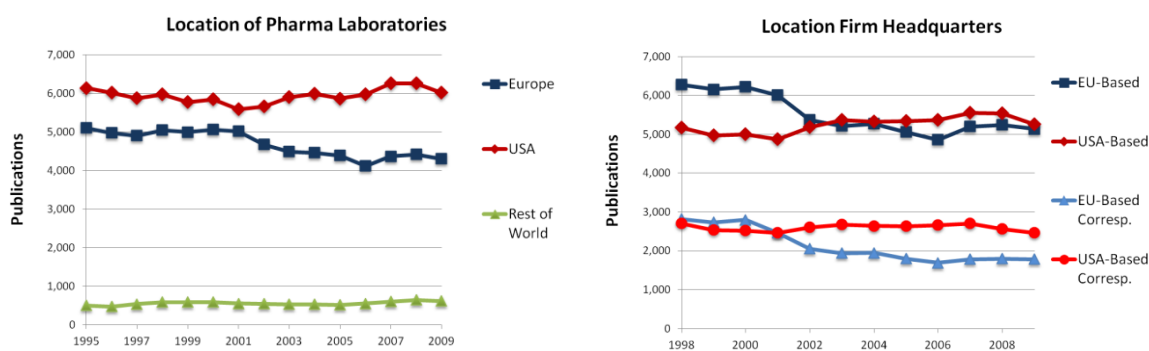

**Figure 6. Publications by location of pharmaceutical R&D laboratories (left). Publications by locations of firm headquarters (right), for all publications (top) and for those with Big Pharma affiliations for the corresponding author (bottom).**

Although in most of this analysis we adopt a Europe versus USA comparative perspective, a more careful analysis of the data reveals that European pharmaceutical companies are still remarkably national (or bi-national as a results of mergers in the case of AstraZeneca and Sanofi-Aventis) (see Table 3). Outside their home countries, European firms have more publications from US-based labs than all their non-domestic European labs (i.e. Europe excluding the 'home country' of the firm). Such is the extent of the national base for collaborations that when co-authorships are mapped into organisational networks there are striking similarities to the natural geographic distribution of countries, as shown in Figure 9, with Big Pharma playing a notable role spanning the bibliometric equivalent of the 'Atlantic'.

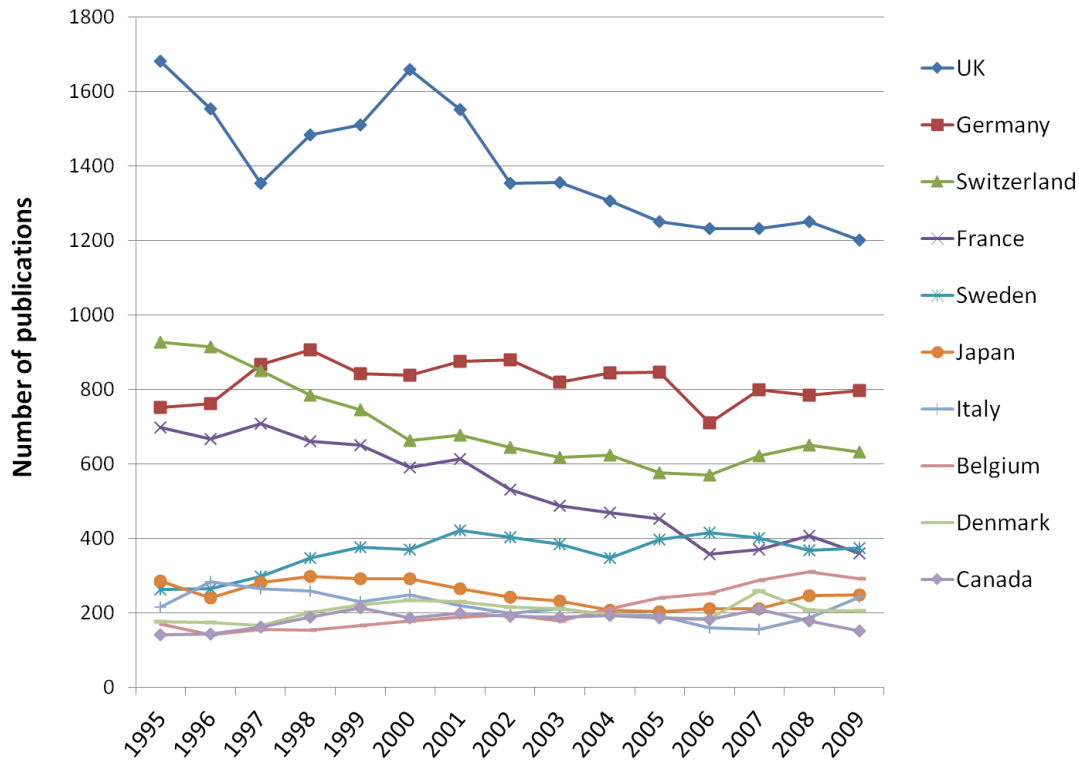

**Figure 7. Location of pharmaceutical R&D laboratories authoring papers**

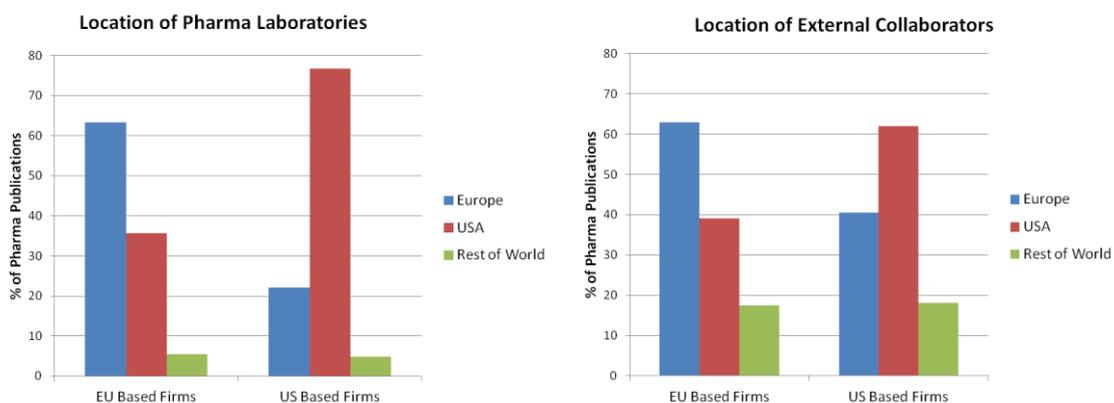

**Figure 8. Location of pharmaceutical R&D laboratories (left) and external collaborators (right) for Big Pharma headquartered in Europe vs. the USA.**

One recurrent argument of why European (or Japanese) pharmaceutical firms locate R&D centres in the US, is that American basic biomedical research is of higher quality than European research (e.g. see (McKelvey et al., 2004))[10]. Given this argument, it seems paradoxical that although USA-based firms have much less R&D in Europe than their EU-based counterparts, both USA and European-based firms have a very similar percentage of their total publications with external collaborators from across the Atlantic (about 40%). Given that the vast majority of these external collaborators are public research organisations, one can argue that in pharmaceutical research European public research is probably not so bad after all. Alternatively it may reflect that the strategic interest of USA-based firms in the European pharmaceutical market is equal to the interest of European-based firms in the USA pharmaceutical market, as local collaborations are needed to prepare drug launches in those markets. Given this interest, scientific activities may well be employed across the Atlantic for marketing reasons, patient availability and proximity to regulators and medical practitioners. The rapid increase in clinical trial activities in Central and Eastern European countries may also be a reason for this.

**Table 3. Comparison between 'Home country' and regional location of research in 1995-2009.**

| | | % of Publications | | | | % Patents[2] |
|---|---|---|---|---|---|---|
| **Firm** | **Home Country** | **Home Country** | **Europe[1]** | **USA** | **Rest of World** | **by US inventors** |
| **USA Firms** | | | | | | |
| **Pfizer** | USA | See USA | 24.2 | 75.6 | 3.4 | 81 |
| **Merck** | USA | See USA | 24.3 | 71.0 | 8.3 | 88 |
| **Eli Lilly** | USA | See USA | 19.9 | 81.7 | 3.9 | n.a. |
| **Johnson & Johnson** | USA | See USA | 33.9 | 65.7 | 4.7 | 86 |
| **Abbott** | USA | See USA | 22.7 | 77.0 | 1.8 | 90 |
| **Bristol-Myers Squibb** | USA | See USA | 7.2 | 93.0 | 2.0 | 90 |
| **Amgen** | USA | See USA | 3.2 | 94.4 | 3.8 | n.a. |
| **Aggregate USA** | | 76.8 | 22.0 | 76.8 | 4.8 | |
| **European Firms** | | | | | | |
| **GlaxoSmithKline** | UK | 44.5 | 15.8 | 43.9 | 2.4 | 47 |
| **Novartis** | Switzerland | 41.1 | 21.6 | 40.2 | 4.4 | 26 |
| **Hoffmann–La Roche** | Switzerland | 24.4 | 17.6 | 50.8 | 10.7 | 46 |
| **AstraZeneca** | UK & Sweden | 77.7 | 3.7 | 19.4 | 3.4 | 19 |
| **Sanofi-Aventis** | Fran. & Germ.[3] | 63.5 | 11.5 | 23.9 | 5.3 | 20 |
| **Bayer** | Germany | 62.2 | 6.8 | 29.2 | 5.7 | n.a. |
| **Novo Nordisk** | Denmark | 87.5 | 4.3 | 9.1 | 2.3 | n.a. |
| **Boehringer Ingelheim** | Germany | 40.8 | 17.2 | 34.9 | 14.7 | n.a. |
| **Aggregate Europe[4]** | | 51.5 | 14.0 | 35.7 | 5.4 | n.a. |
| **Aggregate Europe** | | | 63.3 | 35.7 | 5.4 | n.a. |

Notes: (1) In this table, 'Europe' excludes the home country in the case of European firms. (2) Source: Pammolli et al. (2011, p.434). Based on location of pharmaceutical firms' inventors on patents from 1980-2004. (3) The headquarters of Sanofi-Aventis are in France, but during the period under study one of its antecedent

---
[10] There is a parallel discussion on the higher quality of US research in relation to other industries (Dosi et al., 2006).

companies, Hoechst, had its headquarters in Germany. (4) This European aggregate is based on the sum of European firms. It includes double counting (estimated at 2.5%) due to collaboration between firms.

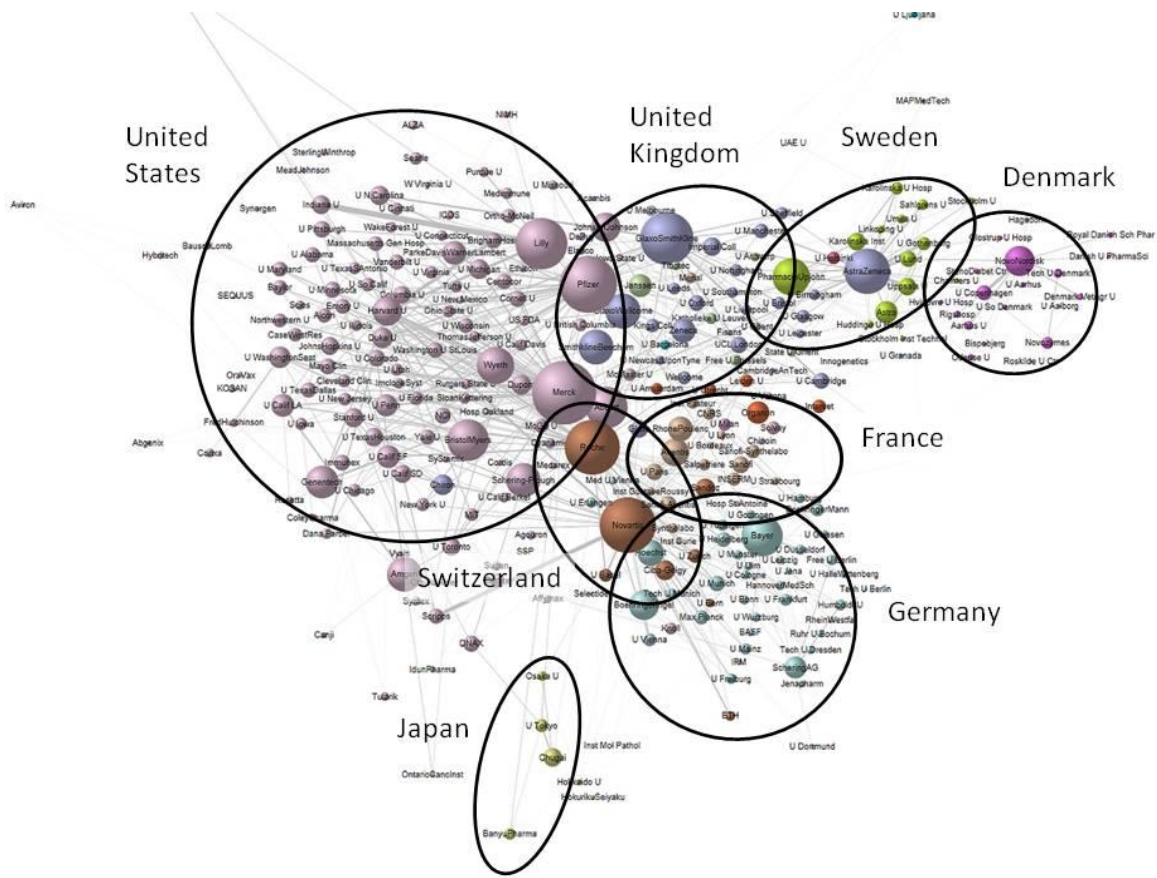

**Figure 9. Collaboration network of pharmaceutical firms.** Different countries are distinguished by shade colour. Ovals only indicate approximate position of national clusters.

**4.3.2 Limited globalisation – so far.**

If there is a decline of European R&D associated with off-shoring and outsourcing of R&D, is there evidence of an increase in publications of R&D in emerging countries? The answer, for now, is certainly not a resounding yes. The percentage of Big Pharma publications outside of USA and Europe has remained stable and low, increasing only from about 500 to about 650 publications per year over the period 1995-2009, although in relative terms these publications have increased from 4% to 6% of Big Pharma's total. Most of these publications (~60-70%) originate in advanced economies such as Japan and Canada. Following investments in R&D centres ("Evolving R&D for emerging markets," 2010) countries such as Singapore, India and China show rapid growth in publications, but start from a very low base. The absolute number of publications emerging from Big Pharma's new eastern R&D laboratories is still very low. As shown in Figure 10, with the exception of Singapore, all other countries still have less than 20 publications/year on average, compared with ~200 in Italy or ~600 in Switzerland.[11]

In contrast, emerging countries are much more important contributors to publication activity when focusing on the external collaborators of Big Pharma. Whereas in 1998, only 10% of Big Pharma collaborators were from outside of Europe or the USA, this figure increased to 15% by 2009. This growth is partly due to the growing importance of universities in developed economies such as Canada and Australia (35% and 17% of 'rest of the world' collaborative publications in 2009, respectively). However, part of the increase is also due to the rapid increase in the collaborations between Big Pharma and public research organisations in countries such as China (157 publications in 2009), Brazil (98), South Korea (75), India (57) and Singapore (55). These collaborations are mainly between local public research organisations and R&D labs of Big Pharma in other countries. As shown in Table 4, co-location of local research organisations and local Big Pharma R&D labs does not appear to be the driver behind the increase of Big Pharma's collaborative publications by emerging countries as these account for only the minority of collaborations that Big Pharma have with authors in the countries listed.

---

[11] The numbers are so low that one wonders if research carried out in R&D laboratories in emerging countries is published with affiliations of the headquarter laboratories. Such practice has been observed in patents.

A more plausible explanation for the increasing collaboration of pharma with emerging economies is the rapid globalization of clinical research activities (Thiers et al., 2008; Petryna, 2009). The conduct of large scale clinical trials requires the involvement of many patients and this renders a division of labour between those researchers that manage trials, design the study and analyze the data, and (clinical) investigators that enrol patients for data analysis. It is the latter's activity that is increasingly conducted in emerging economies and in this case relations between pharmaceutical companies and clinical investigators from these countries are often mediated by a third party such as a Contract Research Organizations (CROs). The increase in external collaborators in emerging economies may therefore be a reflection of the increase of CROs. It is in this context likely that the observed publication activity of emerging economies is an underestimation of actual activity, because authorship for publication mainly accrues to researchers with scientific leadership and less often to the researchers that are actually engaged with patients in this type of clinical work (Hoekman et al., 2012).

In summary, publication data raises questions about the extent to which the opening of R&D centres in emerging economies such as India or China constitute a globalisation of pharmaceutical R&D –in the sense of making these new centres 'competitors' to those in the developed countries. The small number of publications from Big Pharma labs in these countries is consistent with relatively small units of 70 to 300 researchers in most cases (Ujjual et al., 2011), but also suggests these R&D labs are so far fulfilling other purposes, in particular an adaptation of their products to the emerging markets. These include: (i) research on issues specific to human and physical geographies such as tropical diseases or genetic susceptibilities; (ii) development of low cost drugs to cater for the medical needs in terms of cost and dosage of developing country populations; (iii) establishment of links with CROs conduction clinical trials, as well as broadening the genetic make-up of clinical trial populations (iv) presence in large markets with growing wealth and its associated medical needs, e.g. diabetes, cardiovascular complications ("Evolving R&D for emerging markets," 2010; Ujjual et al., 2011, p. 22).

Since the globalisation of pharmaceutical R&D centres is a recent phenomenon and sustained publication activity takes time, there is the possibility that publication counts do not capture the most up-to-date trends. Singapore, the emerging country with most publications, saw the investment in new pharma labs (Lilly, Novartis, Pfizer) in the early 2000s, whereas most centres in China have only opened since the mid 2000s (Roche in 2004; Pfizer and Sanofi-Aventis in 2005; GSK and AstraZeneca in 2007; Novartis, 2008; Lilly and Johnson and Johnson in 2009; source: ("Evolving R&D for emerging markets," 2010)).

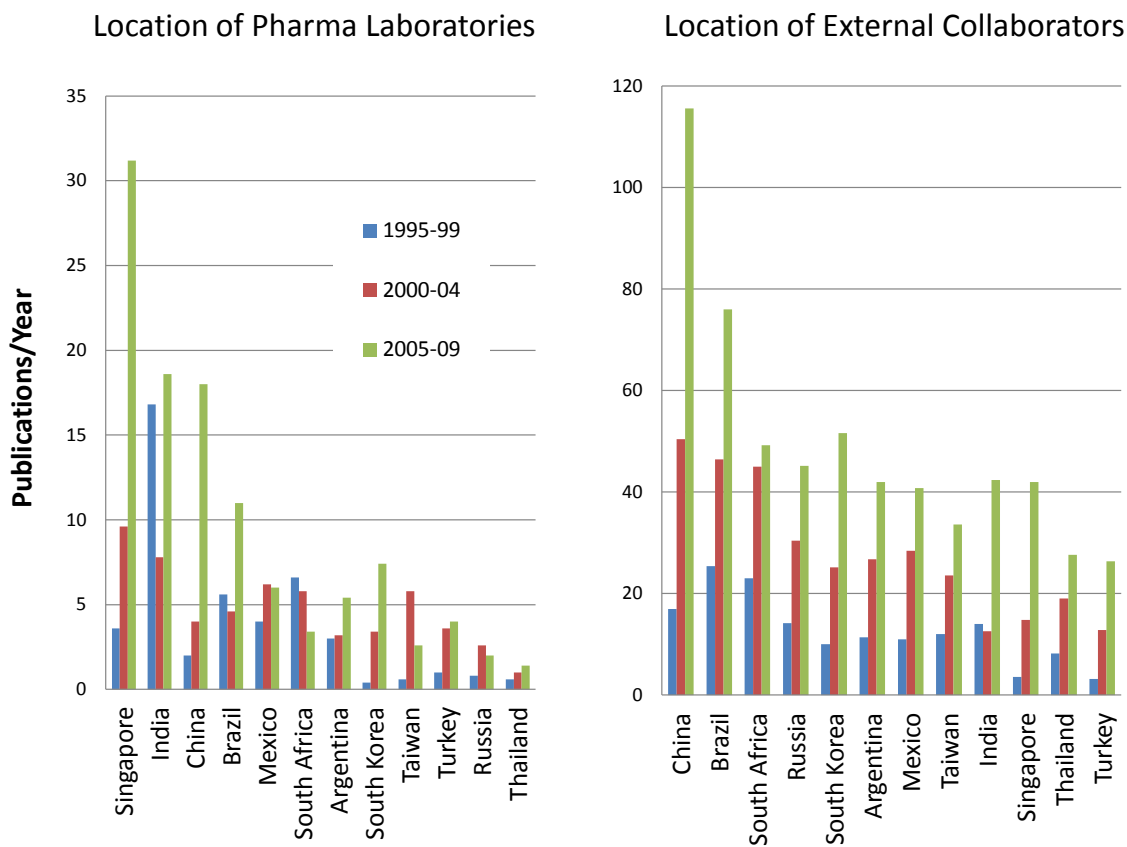

 **Figure 10. Location of Big Pharmas' pharmaceutical R&D laboratories and collaborators in emergent countries.**

**Table 4. Collaborations between local research organisations and local R&D labs of Big Pharma in emerging countries.** The figures indicate the percentage of Big Pharma's publications in a country carried out in collaboration with local external partners.

| Country | % of Big Pharma's collaboration in the country with local external partners | | |
|---|---|---|---|
| | 1995-99 | 2000-04 | 2005-09 |
| India | 19% | 36% | 26% |
| Singapore | 11% | 18% | 19% |
| China | 6% | 4% | 13% |
| South Korea | 4% | 13% | 13% |
| Brazil | 17% | 9% | 13% |
| South Africa | 15% | 9% | 4% |

## 5. Discussion: Causes and consequences of Big Pharma's retreat from Open Science

This paper has analysed the publication output of 15 largest pharmaceutical firms in Europe and the USA, including publications by all acquisition targets of these firms in the period 1995-2009, which represents the core of the traditional pharmaceutical industry – Big Pharma. The analysis suggests that these firms are undergoing a shift away from the open science activities associated with drug discovery and towards a systems integrator role (Munos 2009; Paul et al. 2010; Pammolli, Magazzini, and Riccaboni 2011; Hopkins et al. 2007).

Our empirical data provides some evidence of this shift from three perspectives. In the first place, a diversification of the knowledge base away from the traditional expertise in chemistry and biology towards computation, health services and more clinical fields. This diversification may be characterised crudely as a shift from basic towards clinical research, from research to development, or from bench towards bedside. Such a shift might be resulting from technological opportunities created by information technologies (Nightingale, 2000; Bonaccorsi, 2008), as the outcome of socio-economic pressures in medicine towards practices that are more sensitive to the patient's context or individual needs (Amir-Aslani and Mangematin, 2010) and/or of management strategies to reduce the financial risk associated with drug discovery (Munos and Chin, 2011)[12].

Second, to access these new areas of knowledge we observe an increase in external collaborations and of external leadership of co-authored papers, suggesting a tendency to outsource, in agreement with industry reports of increasing externalisation of R&D efforts (Baum, 2010; Hirschler and Kelland, 2010). Such trends are consistent with the view that the locus of innovation is shifting from in-house R&D to small firms and public organisations (Munos, 2009; Kneller, 2010) or to the organisational network (Powell et al., 1996), in agreement with the Open Innovation model (Chesbrough, 2003). However these expanding collaborations are increasingly intellectually led by external collaborators rather than by big pharma themselves. Whether this has any implications for big pharma's ability to capture value from these relationships remains to be explored.

Third, from a geographic perspective, we find a more pronounced decline in publications by Europe-based big pharma (including their USA labs) as well as a decline in total output of their European labs (including those owned by US firms), in comparison to all big pharma labs located in the US or all labs owned by USA firms. The relative decline in European laboratories is not uniform, with Germany resisting the trend, while the UK, France and Switzerland succumb most. Sweden, Denmark and Belgium buck the trend entirely. The shift away from undertaking internal R&D in Europe does not detract from the importance of European collaborators for Big Pharma – with European collaborators apparently as important to USA-based Big Pharma as US collaborators are to Europe-based Big Pharma.

Finally, while publications by Big Pharma in emergent countries have sharply increased, from a low base over the period studied, these remain extremely low compared with those in developed countries. Publishing activity in these R&D centres in developing countries have so far a small 'research' component and a larger 'development' component, possibly associated with adaptation to local growing markets and coordination of clinical research that is outsourced to local CROs or public centres (Hoekman et al., 2012).

The overarching tendency according to which we should judge these trends is one of a slow decline of the total number of publications by Big Pharma, in stark contrast with the inflationary tendencies of most bibliometric indicators in the period (Persson et al., 2004). This confirms earlier suggestions of a publication decline from corporate laboratories, inferred from a shorter period (1996-2001) and only for certain collaborative modes (Tijssen, 2004). Based on our empirical data we can merely argue that there seems to be a shift away from an Open Science strategy in Big Pharma, which might be due to either a decrease of R&D performed in-house by pharma (possibly due to increased R&D outsourcing), or a decrease in the propensity to publish. As carefully argued by Tijssen (Tijssen, 2004, pp. 726–727), one cannot rule out the possibility that Big Pharma is conducting the same or greater amount of R&D, but just publishing less –for example, because academic collaborators refrain from mentioning industrial involvement or because of fears of knowledge leaking out before patenting. Or it could also be that the retreat from publishing activities is due to a combination of increasing costs in the business of medical writing, heightened awareness by medical journals of the 'marketing' function of pharma articles and/or in the wake of scandals following ethically dubious practices such as ghost writing (Smith, 2005; Sismondo, 2009).

---

[12] See (Pammolli et al., 2011) for an opposite view on risk taking.

However, industry analysts report of have observed R&D laboratories closures (LaMattina, 2011), decrease of R&D expenditure in comparison to sales (from the traditional 15-20% to ~10% (Petsko, 2011, p. 3)), outsourcing of R&D (Baum, 2010; Hirschler and Kelland, 2010), relative reduction of research in comparison to development (Jensen, 2010) and a drop by half in number of new pharma-biotech alliances (Ratner, 2012). This strongly suggests that pharma's in-house R&D efforts are decreasing significantly. In fact, the decline in publication trends shown in Figure 1 might be a gross underestimate of the degree of in-house R&D reduction, because of increasing co-authorship trends observed during the period.

In any case, either of these (perhaps complementary) interpretations tells of a decline in the importance of Open Science consistent with the idea that pharmaceutical companies are increasingly 'network integrators' instead of the prime locus of drug discovery (Subramanian et al., 2011). As the chief executive of GlaxoSmithKline explains:

> 'Big Pharma players can no longer hope to generate the absolutely best science in all areas on their own. Therefore, rebuilding the R&D Engine in Big Pharma standard operating procedure should be to decide on a scientific bet (for example, kinases in oncology), shop around among all the external players that are pursuing such research, and establish a contractual relationship with the best.' (Garnier, 2008, pp. 75–76)

In doing so, Big Pharma is following other high-tech sectors, such as various firms in semiconductors and information technologies in the retreat from corporate R&D (Tijssen, 2004; The Economist, 2007) Instead of being the R&D engine, Big Pharma, as network integrator, is possibly taking on the role of financier, regulatory liaison, lobbyists and salesman for publicly funded medical research because they have the resources, expertise and social capital to navigate the increasingly complex environment of health business, regulations and politics.

The move away from Open Science (sharing of knowledge through scientific conferences and publications) is compatible and consistent with the increasing importance of Open Innovation (increased sharing of knowledge –but not necessarily in the public domain). More specifically, Big Pharma is not merely retreating from publication activities but in doing so it is likely to substitute more general dissemination of research findings in publications for more exclusive direct sharing of knowledge with collaboration partners. Hence, the reduction in publication activities - next to R&D cuts and lab closures – is indicative of a shift in Big Pharma's knowledge sharing and dissemination strategies.

Putting this view in a broader historical perspective, one can interpret the retreat of Big Pharma from Open Science, as the recognition that science (unlike specific technological capabilities) was never a core competence of pharmaceutical firms (Pisano, 2006) and that publication activity required a lot of effort, often without generating the sort of value expected by shareholders. When there are alternatives ways to share knowledge with partners, e.g. via Open Innovation agreements, these may be attractive. Indeed an associated benefit of this process is that Big Pharma can shield itself from scrutiny in the public domain by shifting and distributing risk exposure to public research organisations and small biotech firms.

Whether the retreat from R&D and the focus on system integration is a desirable development depends on the belief in the capacities of big pharma to coordinate and integrate these activities for the public good. At this stage, one can only speculate on the implications of Big Pharma's retreat from Open Science. In a linear view of innovation where high-quality scientific activity results in new compounds, Big Pharma's lack of investment in in-house science (or at least in genuine high risk research) can be seen as the cause of its R&D productivity crisis (Munos and Chin, 2011). However one can easily reverse this argument by stating that R&D cuts are just the consequence of the crisis: higher R&D costs and expiring patents (or anticipation of these events) lead to lack of investment which results in a shortage of capital to keep R&D.[13]

Any discussion of the causality chains in Big Pharma's apparent structural transformation will be complex and is beyond the scope of this study. An understanding of the industry evolution would require an analysis not only of corporate R&D activities, but of the interplay of financial markets, emerging economies, increased regulatory stringency, demographic changes resulting in new types of diseases to be addressed, technologies used in drug discovery, and changes in health provision triggered by welfare cuts. We nevertheless speculatively advance two debates where the retreat of Big Pharma from Open Science will figure centrally.

The first debate concerns the capability of Big Pharma to carry out the role of system integrator in light of a reduction of its scientific base. It has been argued that over time science-based knowledge is becoming more important in drug discovery, and that firms therefore need to develop *absorptive capacity* by interacting with academia and small biotechs in order to acquire this knowledge (Cockburn and Henderson, 1998). This begs the question how Big Pharma will be able to screen and select the areas in which it will invest if it keeps reducing its in-house research. The apparent answer appears to be for academia to step in by doing 'translational research' and setting up private-public centres that facilitate Open Innovation'(West and Nightingale, 2009; Collins, 2011). The viability of this model is however unproven.

The second debate concerns how the redistribution of research efforts will or should affect the redistribution of the social benefits and economic returns derived from drugs. Some observers have seriously questioned the economic returns enjoyed by Big Pharma in the last 20 years (Angell, 2004; Mazzucato, 2011, p. 96), in particular given the crucial contributions already made by public research (McMillan et al., 2000) and the forthcoming increases (Collins, 2011). Reuters' analysts have put it eloquently (Hirschler and Kelland, 2010, p. 9):

> 'Drug companies have long promoted the idea that they pursue new drugs for the good of humanity; it's an argument Big Pharma regularly uses to justify the huge profits it makes. High returns, the industry argues, can be ploughed back into research on the next medical breakthrough. If Big Pharma is not doing the research itself, will the big margins be harder to defend?'

The use of the term defend here is ambiguous and could be interpreted as morally defend, or even defend against appropriation by partner organisations. The latter point features extensively in debates on the commercialization of scientific research which question profitability and raise the prospect of alternative forms of R&D network structures (Vallas and Kleinman, 2008).

In conclusion, the analysis of Big Pharma's publications over a 15 years period has provided us with a window on what may be early signs of a transformation in the pharmaceutical industry, driven by a 'perfect storm' of pressures to change (Staton, 2011). Multidimensional studies weaving observed change in the knowledge base with financial, regulatory and social trends will be needed to put the findings of this paper in perspective and to think through future scenarios in health provision (Crommelin et al., 2010).

---

[13] We thank Ed Steinmueller for this point.

**Acknowledgments**

We thank Mike O'Neill, Luigi Orsenigo and Sergio Sismondo for comments. Ismael Rafols and Alice O'Hare were funded by the EU FP7 project FRIDA (Grant 225546, http://www.fridaproject.eu) and the US NSF (Award no. 0830207, http://idr.gatech.edu/). Michael Hopkins and Josh Siepel were supported by ESRC grant RES-360-25-0076. Jarno Hoekman has been supported by the Netherlands Organisation for Scientific Research (NWO) under the VIDI programme, number 452-06-005. Antonio Perianes-Rodríguez conducted his research at the University of Sussex as awardee of José Castillejo grant, JC2010-0042, funded by the Spanish Ministry of Education. The findings and observations contained in this paper are those of the authors and do not necessarily reflect the views of the funders.

**References**

Amir-Aslani, A., Mangematin, V., 2010. The future of drug discovery and development: Shifting emphasis towards personalized medicine. Technological Forecasting and Social Change 77, 203–217.

Angell, M., 2004. The truth about the drug companies. How they deceive us and what to do about it. Random House, New York.

Archibugi, D., Iammarino, S., 1999. The policy implications of the globalisation of innovation. Research Policy 28, 317–336.

Arrowsmith, J., 2012. A decade of change. Nat Rev Drug Discov 11, 17–18.

AstraZeneca, 2010. AstraZeneca Annual Report 2010. AstraZeneca, London.

Baum, A., 2010. Pharmaceuticals: Exit research and create value.

BIS, 2009. [ARCHIVED CONTENT] Policy: Innovation: R&D Scoreboard - BIS [WWW Document]. URL http://webarchive.nationalarchives.gov.uk/20101208170217/http://www.innovation.gov.uk/rd_scoreboard/?p=31

Bonaccorsi, A., 2008. Search regimes and the industrial dynamics of science. Minerva 46, 285–315.

Chesbrough, H., 2003. Open Innovation: The New Imperative for Creating and Profiting from Technology. Harvard University Press, Boston.

Cockburn, I.M., Henderson, R.M., 1998. Absorptive Capacity, Coauthoring Behavior, and the Organization of Research in Drug Discovery. The Journal of Industrial Economics 46, 157–182.

Collins, F.S., 2011. Reengineering Translational Science: The Time Is Right. Science Translational Medicine 3, 90cm17.

Crommelin, D., Stolk, P., Besancon, L., Shah, V., Midha, K., Leufkens, H., 2010. Pharmaceutical sciences in 2020. Nat Rev Drug Discov 9, 99–100.

D'Este, P., 2005. How do firms' knowledge bases affect intra-industry heterogeneity?: An analysis of the Spanish pharmaceutical industry. Research Policy 34, 33–45.

Dasgupta, P., David, P.A., 1994. Toward a new economics of science. Research Policy 23, 487–521.

David, P.A., 1998. Common agency contracting and the emergence of" open science" institutions. The American Economic Review 88, 15–21.

DiMasi, J.A., Faden, L.B., 2011. Competitiveness in follow-on drug R&D: a race or imitation? Nat Rev Drug Discov 10, 23–27.

Dosi, G., Llerena, P., Labini, M.S., 2006. The relationships between science, technologies and their industrial exploitation: An illustration through the myths and realities of the so-called "European Paradox." Research Policy 35, 1450–1464.

van Eck, N., Waltman, L., 2010. Software survey: VOSviewer, a computer program for bibliometric mapping. Scientometrics 84, 523–538.

Evolving R&D for emerging markets., 2010. Nat Rev Drug Discov 9, 417–420.

Friedman, Y., 2010. Location of pharmaceutical innovation: 2000–2009. Nat Rev Drug Discov 9, 835–836.
Garnier, J.-P., 2008. Rebuilding the R&D Engine in Big Pharma. Harvard Business Review 69–76.
Herper, M., 2011. A Decade In Drug Industry Layoffs [WWW Document]. URL http://www.forbes.com/sites/matthewherper/2011/04/13/a-decade-in-drug-industry-layoffs/
Hicks, D., 1995. Published Papers, Tacit Competencies and Corporate Management of the Public/Private Character of Knowledge. Industrial and Corporate Change 4, 401–424.
Hicks, D., Katz, J.S., 1996. Where Is Science Going? Science, Technology, & Human Values 21, 379–406.
Hirschler, B., Kelland, K., 2010. Big Pharma, Small R&D. Reuters, London.
Hoekman, J., Frenken, K., de Zeeuw, D., Lambers-Heerspink, H.J., 2012. The geographical constitution of leadership in globalized clinical trials. mimeo.
Hopkins, M.M., Martin, P.A., Nightingale, P., Kraft, A., Mahdia, S., 2007. The myth of the biotech revolution: An assessment of technological, clinical and organisational change. Research Policy 36.
Howells, J., Gagliardi, D., Malik, K., 2008. The growth and management of R&D outsourcing: evidence from UK pharmaceuticals. R&D Management 38, 205–219.
Hwang, J., Christensen, C.M., 2008. Disruptive Innovation In Health Care Delivery: A Framework For Business-Model Innovation. Health Affairs 27, 1329–1335.
Jensen, D.G., 2010. Little "r," Big "D" --Science Jobs [WWW Document]. URL http://sciencecareers.sciencemag.org/career_magazine/previous_issues/articles/2010_12_17/caredit.a1000122
Jong, S., Slavcheva, K., 2012. To share or not to : Publishing strategies and R&D productivity in science-based industries. In preparation.
Kneller, R., 2010. The importance of new companies for drug discovery: origins of a decade of new drugs. Nat Rev Drug Discov 9, 867–882.
LaMattina, J.L., 2011. The impact of mergers on pharmaceutical R&D. Nature Reviews Drug Discovery 10, 559–560.
Leydesdorff, L., Rotolo, D., Rafols, I., 2012. Bibliometric Perspectives on Medical Innovation using the Medical Subject Headings (MeSH) of PubMed. Journal of the American Society for Information Science and Technology In press.
Leydesdorff, L., Wagner, C., 2009. Is the United States losing ground in science? A global perspective on the world science system. Scientometrics 78, 23–36.
Mazzucato, M., 2011. The entrepreneurial state. Demos, London.
McBride, R., Hollmer, M., 2012. Top 10 pharma layoffs of 2011 [WWW Document]. URL http://www.fiercepharma.com/special-reports/top-10-pharma-layoffs-2011?utm_medium=nl&utm_source=internal
McKelvey, M., Orsenigo, L., Pammolli, F., 2004. Pharmaceutical analized through the lens of a sectoral innovation system, in: Malerba, F. (Ed), Sectoral Innovation Systems. Concepts, Issues and Analyses of Six Major Sectors in Europe. Cambridge University Press, Cambridge, pp. 73–120.
McMillan, G.S., Narin, F., Deeds, D.L., 2000. An analysis of the critical role of public science in innovation: the case of biotechnology. Research Policy 29, 1–8.
Merton, R., 1973. The normative structure of science, in: Press, U. o. C. (Ed), The Sociology of Science. Chicago and London, pp. 268–278 (tbc).
Montori, V.M., Guyatt, G.H., 2008. Progress in Evidence-Based Medicine. JAMA: The Journal of the American Medical Association 300, 1814–1816.
Munos, B.H., 2009. Lessons from 60 years of pharmaceutical innovation. Nature Reviews Drug Discovery 8, 959–968.

Munos, B.H., Chin, W.W., 2011. How to revive breakthrough innovation in the pharmaceutical industry. Science translational medicine 3, 89cm16.
Nightingale, P., 2000. Economies of scale in experimentation: knowledge and technology in pharmaceutical R&D. Industrial and Corporate Change 9, 315–359.
Nightingale, P., Martin, P., 2004. The myth of the biotech revolution. Trends in Biotechnology 22, 564–569.
Pammolli, F., Magazzini, L., Riccaboni, M., 2011. The productivity crisis in pharmaceutical R&D. Nat Rev Drug Discov 10, 428–438.
Patel, P., Arundel, A., Hopkins, M.M., 2008. Sectoral Innovation Systems in Europe: Monitoring, Analysing Trends and Identifying Challenges in Biotechnology Europe Innova. SPRU, University of Sussex, Brighton, UK.
Patel, P., Pavitt, K., 1997. The technological competencies of the world's largest firms: Complex and path-dependent, but not much variety. Research Policy 26, 141–156.
Paul, S.M., Mytelka, D.S., Dunwiddie, C.T., Persinger, C.C., Munos, B.H., Lindborg, S.R., Schacht, A.L., 2010. How to improve R&D productivity: the pharmaceutical industry's grand challenge. Nat Rev Drug Discov 9, 203–214.
Perianes-Rodríguez, A., Olmeda-Gómez, C., Moya-Anegón, F., 2010. Detecting, identifying and visualizing research groups in co-authorship networks. Scientometrics 82, 307–319.
Persson, O., Glänzel, W., Danell, R., 2004. Inflationary bibliometric values: The role of scientific collaboration and the need for relative indicators in evaluative studies. Scientometrics 60, 421–432.
Petryna, A., 2009. When experiments travel: clinical trials and the global search for human subjects. Princeton University Press, Princeton and Oxford.
Petsko, G.A., 2011. Herding CATS. Science Translational Medicine 3, 97cm24.
Pisano, G.P., 2006. Science Business: The Promise, the Reality, and the Future of Biotech. Harvard Business School Press, Boston, MA.
Polidoro, F., Theeke, M., 2011. Getting competition down to a science: the effects of technological competition on firms' scientific publication. Organization Science doi: 10.1287/orsc.1110.0684.
Porter, A.L., Rafols, I., 2009. Is Science Becoming more Interdisciplinary? Measuring and Mapping Six Research Fields over Time. Scientometrics 81, 719–745.
Powell, W.W., Koput, K.W., Smith-Doerr, L., 1996. Interoganizational collaboration and the locus of innovation: Networks of learning in biotechnology. Administrative Science Quaterly 41, 116–145.
Rafols, I., Porter, A.L., Leydesdorff, L., 2010. Science Overlay Maps: A New Tool for Research Policy and Library Management. Journal of the American Society for Information Science and Technology 61, 871–1887.
Ratner, M., 2012. Big pharma upheavals cast shadow across biotech sector. Nat Biotech 30, 119–120.
Sanofi-Aventis, 2010. Sanofi-Aventis Annual Report 2010. Sanofi-Aventis Groupe, France.
Sismondo, S., 2007. Ghost management: how much of the medical literature is shaped behind the scenes by the pharmaceutical industry? PLoS Medicine 4, e286.
Sismondo, S., 2009. Ghosts in the Machine. Social Studies of Science 39, 171–198.
Smith, R., 2005. Medical Journals Are an Extension of the Marketing Arm of Pharmaceutical Companies. PLoS Med 2, e138.
Staton, T., 2011. Roche chief warns of perfect pharma storm [WWW Document]. Fiercepharma. URL http://www.fiercepharma.com/story/roche-chief-warns-perfect-pharma-storm/2011-12-07
Stephan, P., 1996. The economics of science. Journal of Economic Literature 34, 1199–1235.
Subramanian, R., Toney, J.H., Jayachandran, C., 2011. The evolution of research and development in the pharmaceutical industry: toward the open innovation model – can pharma reinvent itself? International Journal of Business Innovation and Research 5, 63–74.
The Economist, 2007. The rise and fall of corporate R&D: Out of the dusty labs. The Economist.

Thiers, F.A., Sinskey, A.J., Berndt, E.R., 2008. Trends in the globalization of clinical trials. Nat Rev Drug Discov 7, 13–14.
Tijssen, R.J.W., 2004. Is the commercialisation of scientific research affecting the production of public knowledge?: Global trends in the output of corporate research articles. Research Policy 33, 709–733.
Tijssen, R.J.W., 2009. Internationalisation of pharmaceutical R&D: how globalised are Europe's largest multinational companies? Technology Analysis & Strategic Management 21, 859–879.
Timmermans, S., Berg, M., 2003. The practice of medical technology. Sociology of Health & Illness 25, 97–114.
Ujjual, V., Patel, P., Krishnan, R.T., Keshavamurthy, S., Hsiao, R.-L., Zhao, F.Y., 2011. Management and Organisation of Knowledge Creation in Emerging Markets: a Perspective from Subsidiaries of EU MNEs. SPRU Electronic Working Papers 192.
Vallas, S.P., Kleinman, D.L., 2008. Contradiction, convergence and the knowledge economy: the confluence of academic and commercial biotechnology. Socio-Economic Review 6, 283 –311.
West, W., Nightingale, P., 2009. Organizing for innovation: towards successful translational research. Trends in Biotechnology 27, 558–561.
Wuchty, S., Jones, B.F., Uzzi, B., 2007. The Increasing Dominance of Teams in Production of Knowledge. Science 316, 1036 –1039.
von Zedtwitz, M., Gassmann, O., 2002. Market versus technology drive in R&D internationalization: four different patterns of managing research and development. Research Policy 31, 569–588.

**Appendix 1: List of complementary files**

Available at: www.interdisciplinaryscience.net/pharma
- Complementary File 1: List of firm's subsidiaries, acquisitions and mergers
- Complementary File 2: Publication trends by firm
- Complementary File 3: Web of Science Categories (WCs) of publication
- Complementary File 4: Specialisation science maps: aggregate and by firm
- Complementary File 5: Publication and collaboration trends by region and country

The authors are willing to share further materials upon request.

**Appendix2: Methods for the generation of scientometric maps**

Two types of maps were created to visualize the knowledge base of pharmaceutical firms and one type for the collaboration networks. The first type (shown in Figures 3 and 4) was an overlay map of global pharma publications following the methodology detailed in Rafols et al. (2010a). An updated map based on 2009 version of the Global Map of Science was used to map category cross-citations.[14] The second type (used for 2 analyses) mapped the 191 most cited journals by the pharmaceutical firms in the period 1995-2009 (shown in Complementary File 4). All citations received by these 191 journals were downloaded from the WoS. This data was used to create the cosine similarity measure between journals, which allowed location of the papers in a network according to *Pajek*'s Kamada-Kawai algorithm. Factor analysis was applied to the resulting matrix to attribute each journal to a scientific speciality (e.g. Oncology) (Leydesdorff & Rafols, 2009; Perianes-Rodríguez et al., 2010).

As described in the following section, this journal map was used to make two overlays for specialization of given pharmaceutical firms. The third type of map is a co-authorship network of the pharmaceutical firms and the top 50 collaborations of each of them (which have major overlaps), leading to a total of 286 organisations. The standard VOSviewer similarity measure was used to generate the spatial coordinates to set the relative position of organisations. Finally, the network was created in Pajek combining the matrix of co-authorship (edges) and the coordinates imported from VOSviewer (nodes).

---

[14] This basemap is freely available at http://www.leydesdorff.net/overlaytoolkit/.